\documentclass[12pt,letterpaper]{article}
\usepackage{amsmath}
\usepackage{amssymb}
\usepackage{cite}
\usepackage{dcolumn} 
\usepackage{setspace}
\usepackage{array}
\usepackage{graphicx}
\def\brp{{\mathbf{r}^{\prime}}}

\def\br{{\mathbf{r}}}

\def\bR{{\mathbf{R}}}

\def\d{{\mathrm{d}}}
\def\rhor{{\rho({\bf r})}}

\def\rhoi{{\rho_I}}

\def\rhoir{{\rho_I({\bf r})}}

\def\rhojrp{{\rho_J({\bf r}^{\prime})}}
\def\sumi{{\sum_I^{N_S}}}

\newcommand{\eqn}[1]{\mbox{Eq.\hspace{1pt}(\ref{#1})}}

\newcommand{\pot}[1]{v_{\rm #1}}

\begin{document}

\begin{center}
\vspace*{1cm}
{\large\bf Calculating Hyperfine Couplings in Large Ionic Crystals Containing Hundreds of QM Atoms: Subsystem DFT is the Key}\\[3ex]

{\large Ruslan Kevorkyants$^a$\footnote[1]{E-mail: m.pavanello@rutgers.edu, ruslan.kevorkyants@gmail.com\\ \hspace*{6mm}Phone: \hspace*{1mm}973-353-5041\\ \hspace*{6mm}Fax: \hspace*{5mm}973-353-1264}, Xiqiao Wang$^b$, David M. Close$^b$,\\ and Michele Pavanello$^{a*}$}\\[2ex]
$^a$Department of Chemistry, Rutgers University, Newark, NJ 07102, USA\\
$^b$Department of Physics, East Tennessee State University, TN 37614, USA
\end{center}

\vfill

\begin{tabbing}
Date:   \quad\= \today \\
Status: \> Submitted to J. Phys. Chem. B \\
\end{tabbing}
\newpage
\begin{abstract}
We present an application of the linear scaling Frozen Density Embedding (FDE) formulation of subsystem DFT to the calculation of isotropic hyperfine coupling constants (hfccs) of atoms belonging to a guanine radical cation embedded in a guanine hydrochloride monohydrate crystal. The model systems considered range from an isolated guanine to a 15,000 atom QM/MM cluster where the QM region is comprised of 36 protonated guanine cations, 36 chlorine anions and 42 water molecules.

Our calculations show that the embedding effects of the surrounding crystal cannot be reproduced neither by small model systems nor by a pure QM/MM procedure. Instead, a large QM region is needed to fully capture the complicated nature of the embedding effects 
in this system.

The unprecedented system size for a relativistic all-electron isotropic hfccs calculation can be approached in this work because the local nature of the electronic structure of the organic crystals considered is fully captured by the FDE approach.

\end{abstract}
\newpage
\section{Introduction and Background}
Electron Spin Resonance (ESR) spectroscopy is a powerful tool for studying structure and reactivity of radicals \cite{mcon1957,lund2011}. Despite their rather specific features, such as high reactivity and thus short life-time, radicals occur in nature often as intermediates in chemical reactions and play important roles in various chemical, biochemical, and photophysical processes. Radicals, however, can result from irradiation of matter, a process particularly important for living organisms. It is known that irradiation by X-ray causes DNA damage giving rise to a vast variety of organic radicals. Although this enables the use of ESR spectroscopy, studies of irradiated DNA molecules 
are challenging. In order to facilitate these investigations, ESR spectra of simpler systems, e.g.\ X-ray irradiated crystals or solutions of DNA constituents, are often 
considered. Even then, the assignment of ESR spectra to a particular chemical substance might still be a challenge \cite{lund2011}. It is therefore desirable to have access to complementary tools suitable for investigations of radicals.

Alongside with ESR spectroscopy, computational methods provide a mean of addressing properties of radicals avoiding costly experiments. Theoretically, however, the most dauting task is the determination of the electronic structure of a molecular system. 
Presently, the Kohn-Sham Density Functional Theory (KS-DFT) approach \cite{kohn1965} 
is the method of choice for the determination of electronic structures due to a good balance between computational complexity and accuracy. Starting in the 90s, DFT calculation of ESR observables became possible \cite{erik1994}.

Practical implementations of the KS-DFT approach scale computationally as $O(N^3)$, with $N$ being number of electrons. In the past decade many approximations \cite{scus1999,goed1999,bowl2012} could reduce KS-DFT computational complexity particularly for spatially extended molecules. However, the calculation of most realistic, fully-solvated systems is still largely prohibitive 
\cite{carl2002,burk2012}.

One way of reducing computational complexity of KS-DFT is to use density partitioning techniques which are based on the idea of partitioning of the total electron density of a system into subsystem contributions. Computational complexity is thereby drastically reduced \cite{gord1972,kim1974}. Two successful applications of subsystem DFT already took place in the 80s\cite{sen1986} and early 90s \cite{cort1991}, however, it became more widely known after the publication of a paper by Wesolowski and Warshel in 1993 \cite{weso1993}. Subsystem DFT nowadays is being developed by many research groups worldwide \cite{silv2012,hu2012,pava2013a,gome2012,hofe2012,lari2012,good2011a,neug2010a,iann2006}, and its successful applications are reported for ground state properties, such as analysis of electron densities \cite{fux2008}, and spin densities 
\cite{solo2012} and for calculations of charge transfer parameters 
\cite{pava2013a,pava2011b}; as well as for excited state properties 
\cite{tecm2012,koni2012b,koni2011,neug2010a,bulo2008,jaco2006b}. In the next section we briefly outline the theoretical grounds of one such formulations of subsystem DFT that we find particularly suitable for treating large ionic molecular assemblies, the Frozen Density Embedding (FDE).
\subsection{Frozen Density Embedding formulation of subsystem DFT}
KS-DFT can be summarized by the following equation, the KS equation, in canonical form,
\begin{equation}
\label{ks1}
\left[ -\frac{1}{2} \nabla^2 + \pot{eff}(\br) \right] \phi_k(\br) = \varepsilon_k  \phi_k(\br),
\end{equation}
where $\pot{eff}$ is the effective potential that the one-particle KS orbitals, $\phi_k$, experience, and $\varepsilon_k$ are the KS orbital energies. The spin labels have been omitted for sake of clarity. The electron density for singlets is simply $\rhor=2\sum_i^{\rm occ} \left| \phi_i(\br)\right|^2$.

The effective potential, $\pot{eff}$, is given by
\begin{equation}
\label{ks2}
\pot{eff}(\br)=\pot{eN}(\br)+\pot{Coul}(\br)+\pot{xc}(\br),
\end{equation}
where $\pot{eN}$ is the electron--nucleus attraction potential, $\pot{Coul}$ the Hartree potential, and $\pot{xc}$ the exchange--correlation (XC) potential\cite{kohn1965}. 

Subsystem DFT, instead, is based on the idea that a molecular system can be more easily approached if it is partitioned into many smaller subsystems. In mathematical terms, this is done by partitioning the electron density as follows \cite{sen1986,cort1991}
\begin{equation}
\label{fde1}
\rhor=\sumi \rhoir,
\end{equation}
with $N_S$ being the total number of subsystems. 

Self-consistent solution of the following coupled KS-like equations (also called KS equations with constrained electron density \cite{weso2006}) yields the set of subsystem KS orbitals, i.e.
\begin{equation}
\label{ks1}
\left[ -\frac{1}{2} \nabla^2 + \pot{eff}^I(\br) \right] \phi^I_k(\br) = \varepsilon^I_k  \phi^I_k(\br),\mathrm{~with~}I=1,\ldots,N_S
\end{equation}
with the effective subsystem potential given by
\begin{equation}
\label{sks}
\pot{eff}^I (\br)=\underbrace{\pot{eN}^I (\br) + \pot{Coul}^I(\br) +\pot{xc}^I (\br)}_{\mathrm{same~as~regular~KS-DFT}} + \pot{emb}^I (\br).
\end{equation}
In FDE \cite{weso1993,weso2006}, the yet unknown potential above, $\pot{emb}$, is called embedding potential and is given by
\begin{align}
\label{emb}
\nonumber
\pot{emb}^I (\br)&=\sum_{J\neq I}^{N_S} \left[ \int \frac{\rhojrp}{|\br-\brp|} \d\brp - \sum_{\alpha\in J} \frac{Z_\alpha}{|\br-\bR_{\alpha}|} \right] + \\
&+\frac{\delta T_{\rm s}[\rho]}{\delta\rhor}-\frac{\delta T_{\rm s}[\rhoi]}{\delta\rhoir}+\frac{\delta E_{\rm xc}[\rho]}{\delta\rhor}-\frac{\delta E_{\rm xc}[\rhoi]}{\delta\rhoir}.
\end{align}

where $T_{\rm s}$, $E_{\rm xc}$ and $Z_\alpha$ are the kinetic and exchange-correlation energy functionals and the nuclear charge respectively.

The density of the supersystem is found using \eqn{fde1} and \eqn{ks1} for singlets $\rhor=2\sumi\sum_i^{{\rm occ}_I} \left| \phi^I_i(\br)\right|^2$. The above equations have not explicitly taken into account spin, however the full subsystem local spin density approximation equations can be retrieved elsewhere \cite{solo2012,good2011a}.

\subsection{Hyperfine coupling calculations with KS-DFT and FDE}
It is known that FDE accounts well for solvent effects and produces accurate hyperfine coupling constants of neutral radicals in solution \cite{neug2005f}. In the present work, we show that FDE can produce good EPR parameters also for charged radicals embedded in an ionic environment. These calculations are particularly challenging for KS-DFT. On one hand, it is known that the self-interaction error of KS-DFT affects the electron density of radicals resulting in overdelocalized spin/charge densities. This problem can be overcome by employing density functionals containing high percentage of exact Hartree-Fock exchange such as M06-2X used in the present work. A larger problem, however, arises while treating charged radical species embedded in ionic environment, $e.g.$ ionic crystals. It is known that ionic crystals are stabilized by crystal lattice energy. In addition, within the crystal environment ionizaton potentials of anions and electron affinities of cations are different from their values \textit{in vacuo}. In cluster models, however, both of these effects are not fully taken into account and the delicate equilibrium between the two decides whether the charge separated or the charge recombined system is more stable. This equilibrium is further complicated by the self-interaction error present in local and semilocal density functionals, because this error leads to overdelocalized electron (spin) densities. Recently, it has been shown \cite{solo2012,pava2011b} that FDE can effectively localize both charges and spin. The use of local and semilocal non-additive kinetic energy functionals adds spurious repulsive potential walls between subsystems (ions) \cite{fux2010} that counteract artificial electron density flow occurring in ion-pair systems treated with KS-DFT \cite{pava2011b}. Thus, FDE can be seen as an effective solution that works particularly well for systems comprised of non-bonded charged molecular fragments, such 
as the one we tackle in this work.

The particular choice of guanine radical cation made in this work is dictated from one hand by the availability of accurate ESR and ENDOR experiments 
\cite{close2003,close1993,close1992,close1985,close1987}, and on the other hand by the fact that previous theoretical studies of this irradiated crystal were performed with small cluster models that did not comprehensively took into account the embedding effects of the crystal bulk at a QM level \cite{wetm1998,adhi2006,wang2012}. Thus, in addition to the methodological aspect of this work outlined above, here we provide a definitive answer to the extent of the embedding effect on the hyperfine coupling constants in the irradiated guanine hydrochloride monohydrate crystal.

The paper is constructed in the following way. First, we give a brief description of guanine hydrochloride monohydrate crystal structure. In section \ref{sec:details}, the models used to mimic guanine crystal are presented and the details of computational and methodological aspects are explained. Then, in section \ref{sec:resdisc} we present KS-DFT and FDE isotropic hfccs for a simple crystal model containing one guanine radical cation with/without a water molecule and a chlorine ion. Having concluded that KS-DFT applied to the guanine-chlorine ion-pair system is not suitable for describing the electron density of the crystal system, we also report FDE calculations for two larger models of the crystal consisting of 7 and 36 guanine molecules, respectively. The convergence of the calculated hyperfine coupling constants with the cluster size is carried out using the FDE approach in combination with the QM/MM technique and a model system containingi, overall, over 15,000 atoms.

\section{Details of the calculations\label{sec:details}}

\subsection{The Crystal Structure Considered}
\begin{figure}
\begin{center}
\begin{tabular}{m{7cm}m{5cm}}
\begin{center}(a)\end{center} & \begin{center}(b)\end{center}\\
\includegraphics[width=6.8cm]{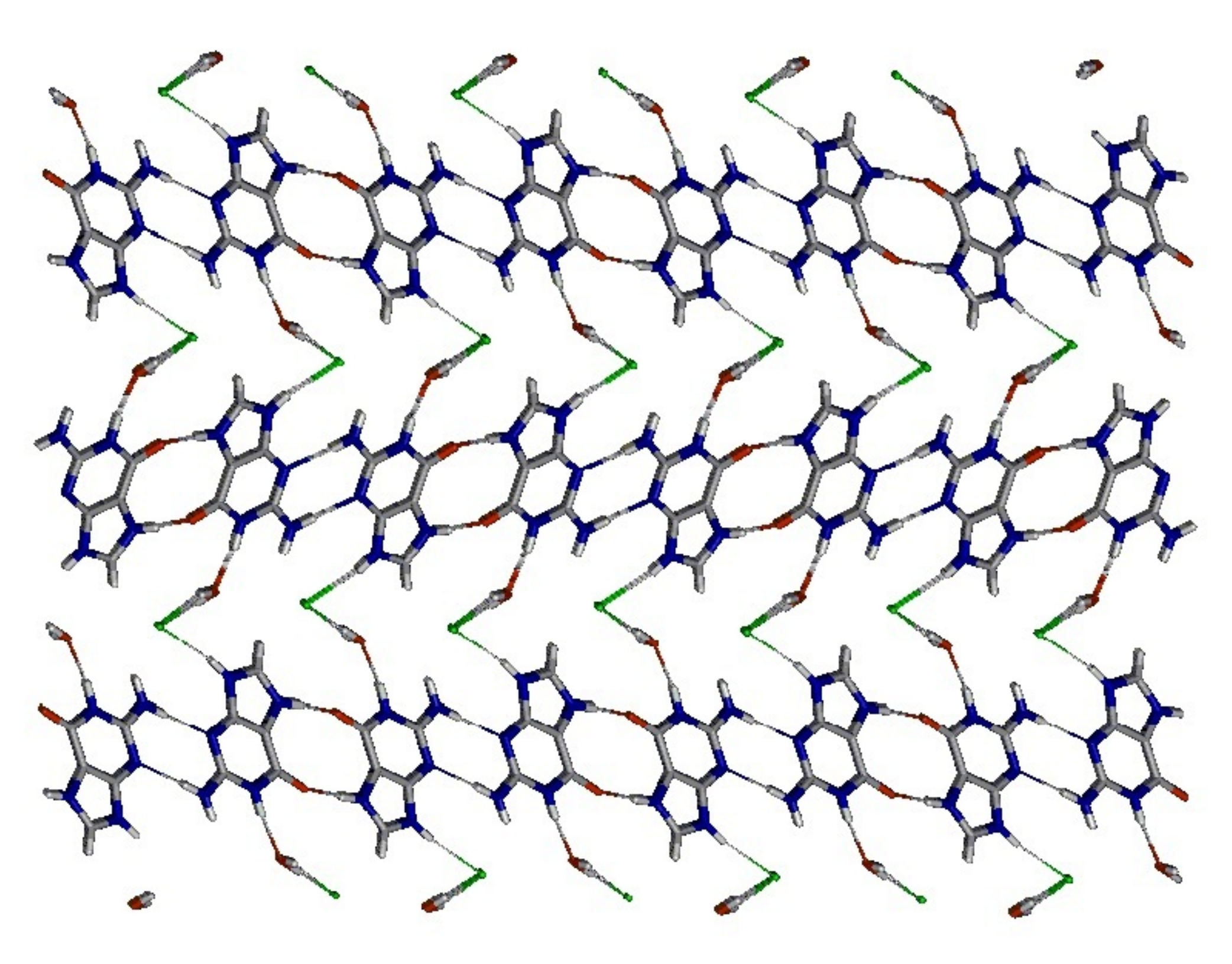}&\includegraphics[width=5cm]{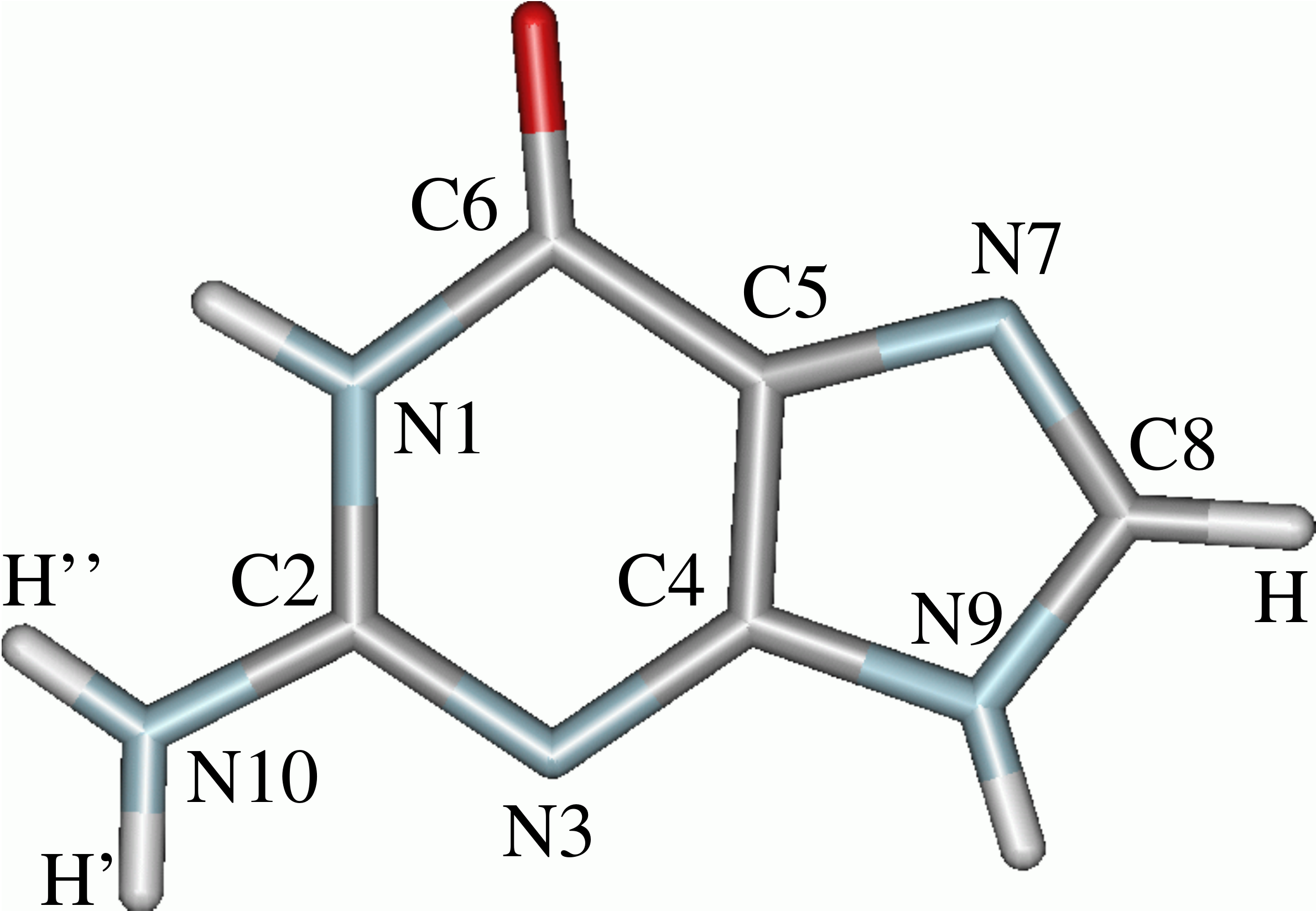}
\end{tabular}
\end{center}
\caption{\label{fig1}Inset (a) Slab of guanine hydrochloride monohydrate crystal; 
inset (b) Guanine molecule deprotonated at the N7 position. 
Carbons, hydrogens, nitrogens, oxygens, and chlorines are shown in 
dark gray, light gray, blue, red, and green, respectively (color online).}
\end{figure}

The high resolution ($R=0.034$) X-Ray diffraction crystal structure of guanine hydrochloride monohydrate (including positions of all hydrogens) is available 
\cite{maix1991} and a portion of it is depicted in inset (a) of Figure \ref{fig1}. 
It has P21/c symmetry and can be described as consisting of blocks of protonated guanine molecules separated from each other by layers of water molecules and chlorine ions. Within such block, the guanine molecules form parallel layers. Finally, within 
a single layer, the guanine molecules are oriented towards each other in alternating face-to-face and back-to-back manner. In both face-to-face and back-to-back orientations, hydrogen bonding takes place. In contrast to typical organic molecular crystals, the crystals of guanine hydrochloride monohydrate are ionic in character. The guanine molecules in the crystal are protonated (the protons originate from hydrochloric acid) and thus carry charge of +1. The protonation site is the nitrogen atom N7 of the imidazole ring, Figure 1, inset (b).

Three models of guanine hydrochloride monohydrate crystal are considered. In all models the atomic positions are the same as in the native crystal described above. The simplest 
model consists of a bare guanine radical cation either with or without water molecule and chlorine ion. The water molecule and the chlorine ion nearest to the pyrimidine ring of a guanine molecule are chosen. Two additional models are considered and consist of 7 and 36 guanine moieties (7G and 36G, hereafter). In contrast to the 7G model system where an exact stoichiometric ratio is obeyed (7 water molecules and 7 chlorine ions are 
included in the model), 42 water molecules are included in the 36G model. All models are 
chosen to be charge neutral.

We only consider one guanine damaged by irradiation in all model systems. The damaged guanine is referred to as the guanine radical cation, hereafter, and it is obtained by one electron oxidation, followed by deprotonation at N7. The structure of guanine radical cation has not been optimized in further calculations which may seem doubtable. We assume, however, that, structures of protonated guanine and guanine radical cation are very similar. 

To assess this assumption we have optimized geometries of both species at PBE/TZ2P level \textit{in vacuo} (see computational details section for explanation). Both structures appeared to be planar. We employed the PBE functional for geometry optimization as this functional yields more reliable geometries than the M06-2X functional \cite{truh2008}.  The largest bond length N7-C1 and bond angle N7-C8-N9 deviations recorded are 0.040 \AA\ and 5.11$^o$, respectively. The M06-2X/TZP isotropic hfccs (MHz; protonated guanine in brackets) are: 9.100 (9.274), 6.461 (7.611), -9.739 (-11.305), -10.958 (-12.651), and -23.754 (-21.298) for N3, N10, H$^{\prime}$, H$^{\prime\prime}$, and H, respectively. The hfccs are very close which, due to their known high sensitivity to structural changes, implies very similar geometries of both species. We do not expect significant structural changes of the guanine radical cation when it is placed within the crystal environment. These considerations are only of qualitative character and do not constitute a thorough analysis of the geometry relaxation of the guanine after deprotonation which will be the focus of a future work.

\subsection{Computational details}

In this work, the implementation of FDE in the Amsterdam Density Functional program \cite{teVelde:2001a,jaco2008b} has been employed. The isotropic hfccs are calculated within the self-consistent Relativistic Zeroth-Order Regular Approximation (ZORA)
\cite{lent1993,vlenthe1998}. Thus, relativistic effects on the subsystem electron density are included in our calculations in a self-consistent fashion.

In all FDE calculations, the systems were split into as many subsystems as non-bonded 
molecules were present in the model. For the two large model systems, 7G and 36G, only 
FDE calculations are performed, $i.e.$ no KS-DFT. In all calculations the basis set dependence of the calculated isotropic hfccs is investigated using TZP and TZ2P Slater-Type Orbital basis sets. Embedding densities and point charges are calculated with TZP basis set. The particular choice of the basis sets made in this work is based upon practical considerations, $e.g.$ it is a good balance between accuracy and computational cost. Additional justification for the use of triple-$\zeta$ quality basis set is grounded on its successfull application in Ref.\ \cite{neug2005f}.

The calculations are all-electron except when explicitly stated. Three freeze-and-thaw FDE cycles are applied to ensure a good degree of self-consistency among all subsystem densities \cite{weso1996b}. The M06-2X \cite{zhao2008} metahybrid XC density functional is employed throughout. This functional contains 54 \% of exact HF exchange which reduces excessive charge and spin delocalization due to self-interaction error, and thus it is preferable over local and semilocal functionals for calculations of radical species. For sake of comparison, in the supplementary materials (see supporting information paragraph), we also report results obtained with the PBE0 functional \cite{ernz1999,admo1999}. The non-additive kinetic energy and exchange--correlation functionals needed to evaluate the embedding potential according to Eq.(6) employed PW91k \cite{lemb1994} and PW91 \cite{perd1991} GGA functionals, respectively.  

All FDE isotropic hfccs calculations are carried out only for the radical cation subsystem. This is justified because our calculations show that the spin density is 
almost completely localized on that guanine, and essentially no inter-subsystem spin transfer occurs. In order to test convergence of the results with respect to the cluster size the bare guanine, 7G and 36G models were encapsulated into a large crystal 
pocket comprised of MM atoms (paritial charges only). In each case the number of atoms totalled 15,119. The contribution from the electrostatic field generated by these embedding atoms (M06-2X/TZP point charges determined for an isolated protonated guanine) into isotropic hfccs was taken into account in a self-consistent manner by including the interaction with the MM charges as a one-electron external potential contribution to the FDE embedding potential.

\section{Results and Discussion\label{sec:resdisc}}
After X-ray irradiation at T=20K, the EPR spectra of guanine crystals indicate the presence of guanine radical cation species \cite{close1985}. The accepted interpretation of these spectra concludes that the X-ray radiation results in one electron oxidation, after which dication deprotonates (presumably at N7 position) thus becoming guanine radical cation. In this section we report and discuss results of FDE calculations of isotropic hfccs of N7 deprotonated and oxidized guanine molecule at different levels of approximations and within various crystal models and compare them with the experiment.

\subsection{Guanine with/without water molecule and chlorine ion}
\begin{figure}
\begin{center}
\includegraphics[width=10cm]{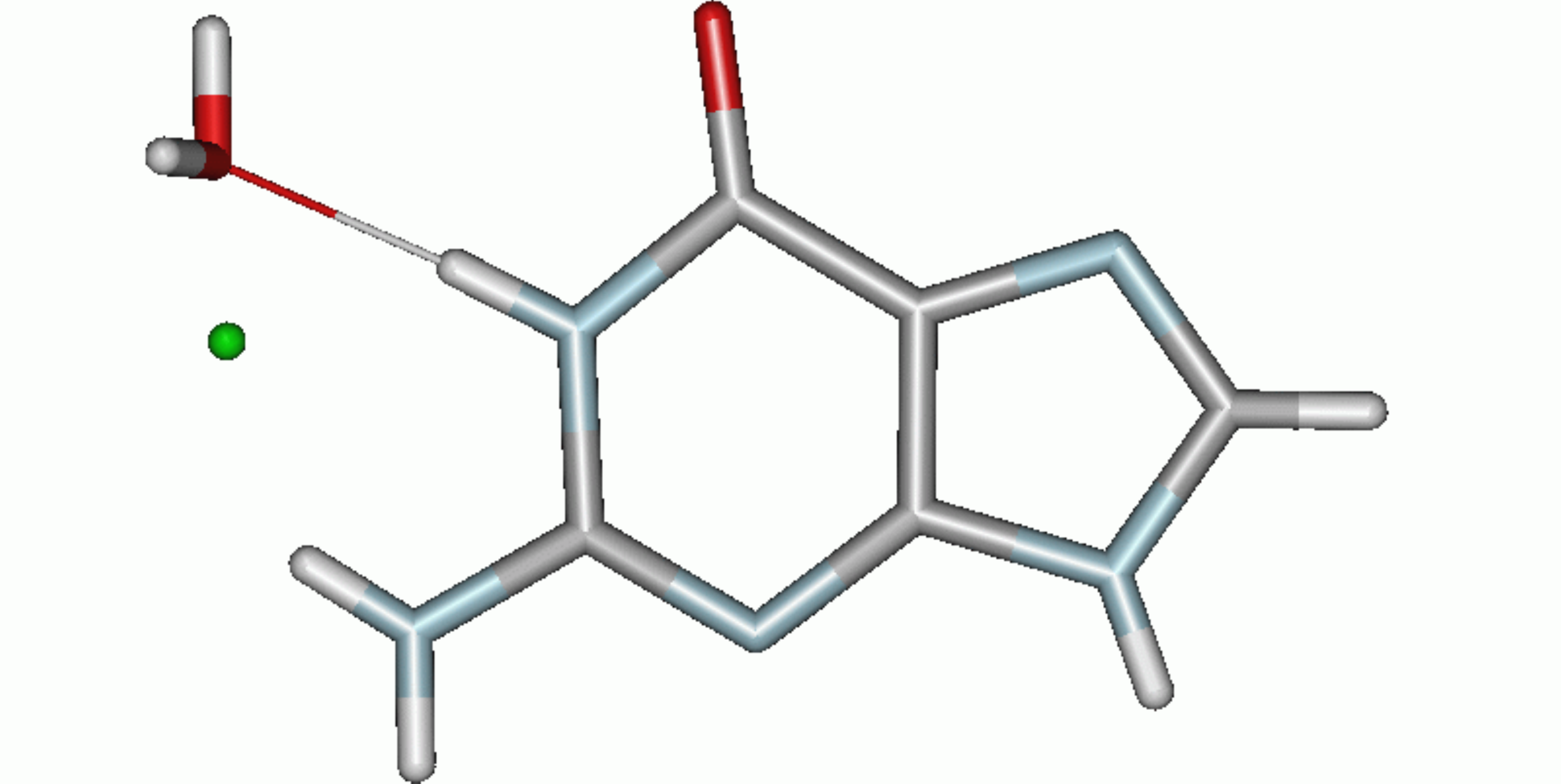}
\end{center}
\caption{\label{fig2}Guanine radical cation with one water molecule and one chlorine ion placed on the side of the pyrimidine ring. Color scheme follows Fig.\ \ref{fig1} (color online).}
\end{figure}

Following the lead of Ref.\ \cite{close1985}, as the EPR spectra of guanine hydrochloride monohydrate crystal were assigned to guanine radical cation, the simplest model to consider is a bare guanine radical cation. The structure of this radical is shown in Figure \ref{fig2}. In order to avoid confusion, we should note that in this figure the bare guanine radical cation is depicted together with one water molecule and one chlorine ion. The structure in Figure \ref{fig2} is the building block of all model systems considered and is shared among all calculations for this model. The calculated isotropic hfccs for bare guanine radical cation are listed in Table \ref{t1} together with the experimental benchmark values.

Despite the simplicity of the bare guanine model, the isotropic hfccs of the non-equivalent atoms qualitatively agree with the experimental values. However, we notice a considerable underestimation of their absolute values for nitrogen atoms, and an overestimation for the hydrogen atom of the imidazole ring [atom H in inset (b) of Figure \ref{fig1}].

\begin{table}
\begin{center}
\begin{tabular}{lrrr}
\multicolumn{4}{l}{{\sc Guanine}}\\
\hline
         &           & \multicolumn{2}{c}{{\bf KS-DFT}} \\
Atom & Ref.\ \cite{close1985}$^a$& TZP&TZ2P\\

                          N3                  &16.8    & 11.8   & 12.0  \\
                          N10                 &10.1    &  7.2   &  7.2  \\
                          H$^{\prime}$        &12.1    & -9.6   & -9.4  \\
                          H$^{\prime\prime}$  &12.1    &-10.6   &-10.0  \\
                          H                   &11.1    &-17.1   &-16.7
\end{tabular}
\end{center}
\caption{\label{t1} KS-DFT M06-2X isotropic hfccs (MHz) for bare guanine radical cation calculated with TZP and TZ2P basis sets. hfcc value for H atom in the last row is obtained from Ref.\ \cite{close1987}.}
\end{table}

Aiming at characterizing the crystal embedding effects on the isotropic hfccs, the presence of water and chlorine ions in the native crystal should be taken into account. In this simple model we include, besides the bare guanine, one water molecule and one chlorine ion, both chosen to be the closest ones to the guanine's pyrimidine ring in the native crystal. Three additional sets of isotropic hfccs calculations are performed: bare guanine radical cation with one water molecule, bare guanine with chlorine ion, and bare guanine with both one water molecule and one chlorine ion. Although the models including either water or chlorine ion do not represent the native crystal, presumably they should help us understand the effects of different crystal constituents on the hfccs and thus should be useful for elaborating economic strategies of such calculations in large systems. 

\begin{table}
\begin{center}
\begin{tabular}{lrrrrr}
\multicolumn{6}{l}{{\sc Guanine + water}}\\
\hline
         &           & \multicolumn{2}{c}{{\bf KS-DFT}} & \multicolumn{2}{c}{{\bf FDE}}\\
Atom & Ref.\ \cite{close1985}$^a$& TZP&TZ2P & TZP & TZ2P\\
                 N3                  & 16.8  & 13.6   & 13.8  & 13.7   & 13.9 \\
                 N10                 & 10.1  &  7.7   &  7.7  &  7.9   &  7.9 \\
                 H$^{\prime}$        & 12.1  & -9.8   & -9.6  & -9.9   & -9.7 \\
                 H$^{\prime\prime}$  & 12.1  &-10.8   &-10.3  &-11.0   &-10.5 \\
                 H                   & 11.1  &-16.9   &-16.2  &-16.8   &-16.4
\end{tabular}
\end{center}
\caption{\label{t2} KS-DFT and FDE M06-2X isotropic hfccs (MHz) for guanine radical cation in the presence of one water molecule calculated with TZP and TZ2P basis sets. $^a$ hfcc value for H atom in the last row is obtained from Ref.\ \cite{close1987}.}
\end{table}

\begin{table}
\begin{center}
\begin{tabular}{lrrrrr}
\multicolumn{6}{l}{{\sc Guanine + chlorine}}\\
\hline
         &           & \multicolumn{2}{c}{{\bf KS-DFT}} & \multicolumn{2}{c}{{\bf FDE}}\\
Atom & Ref.\ \cite{close1985}$^a$& TZP&TZ2P & TZP & TZ2P\\
                 N3                  & 16.8  & 0.8   & 0.5   & 17.6   & 17.8 \\
                 N10                 & 10.1  & 1.4   & 1.3   &  9.5   &  9.5 \\
                 H$^{\prime}$        & 12.1  &-1.2   &-0.9   &-11.9   &-11.7 \\
                 H$^{\prime\prime}$  & 12.1  & 0.2   & 0.3   &-12.6   &-12.0 \\
                 H                   & 11.1  &-0.3   &-0.2   &-15.6   &-15.1
\end{tabular}
\end{center}
\caption{\label{t3} KS-DFT and FDE M06-2X isotropic hfccs (MHz) for guanine radical cation in the presence of one chlorine ion calculated with TZP and TZ2P basis sets. $^a$ hfcc value for H atom in the last row is obtained from Ref.\ \cite{close1987}.}
\end{table}

\begin{table}
\begin{center}
\begin{tabular}{lrrrrr}
\multicolumn{6}{l}{{\sc Guanine + water + chlorine}}\\
\hline
         &           & \multicolumn{2}{c}{{\bf KS-DFT}} & \multicolumn{2}{c}{{\bf FDE}}\\
Atom & Ref.\ \cite{close1985}$^a$& TZP&TZ2P & TZP & TZ2P\\
                 N3                  & 16.8  & 0.5   & 0.5   & 18.2   & 18.4 \\
                 N10                 & 10.1  & 0.6   & 0.6   &  9.6   &  9.6 \\
                 H$^{\prime}$        & 12.1  &-0.5   &-0.4   &-12.1   &-11.9 \\
                 H$^{\prime\prime}$  & 12.1  & 0.1   & 0.7   &-12.9   &-12.3 \\
                 H                   & 11.1  &-0.2   &-0.1   &-15.3   &-14.8
\end{tabular}
\end{center}
\caption{\label{t4} KS-DFT and FDE M06-2X isotropic hfccs (MHz) for guanine radical cation in the presence of one water molecule and one chlorine ion calculated with TZP and TZ2P basis sets. $^a$ hfcc value for H atom in the last row is obtained from Ref.\ \cite{close1987}.}
\end{table}

The effect of inclusion of water molecule and/or chlorine ion on the isotropic hfccs is analyzed using both the KS-DFT and FDE approaches, see Tables \ref{t2}--\ref{t4}. The tables show that in both KS-DFT and FDE treatments, the inclusion of a water molecule into the model has rather minor effect. Upon addition of the chlorine ion, however, KS-DFT and FDE hfccs become dramatically different. The FDE results for all three cases resemble those for bare guanine case. It should be noted, however, that, there are two non-equivalent ways to arrange water and chlorine ion in the model. The results of Table 4 correspond to their placement just as depicted in Figure \ref{fig2}, i.e. at the side of the pyrimidine ring. When Cl$^-$ is located at the side of imidazole ring the following hfcc values are obtained by FDE: 10.078 (N3), 5.882 (N10), -8.070 (H$^{\prime}$), -8.962 (H$^{\prime\prime}$), and -18.256 (H). 
In this arrangement the hfccs for the N3 and N10 atoms decrease by 8 and 4 MHz, respectively. The shift can be explained by noticing that the chlorine ion effectively pushes away the excess charge. Thus, when it is placed near the N3 and N10 atoms (imidazole side) it pushes the excess charge away from them (lower spin density on the N-atoms). We quantified this effect by calculating the Mulliken atomic (diagonal) spin density populations for the N3 and N10 atoms which in fact decreases by 0.15 for N3 and by 0.05 for N10 when Cl$^-$ is placed on the imidazole side. 

\begin{figure}
\begin{center}
\begin{tabular}{c}
(a)\\[1.5ex]
\includegraphics[width=6cm]{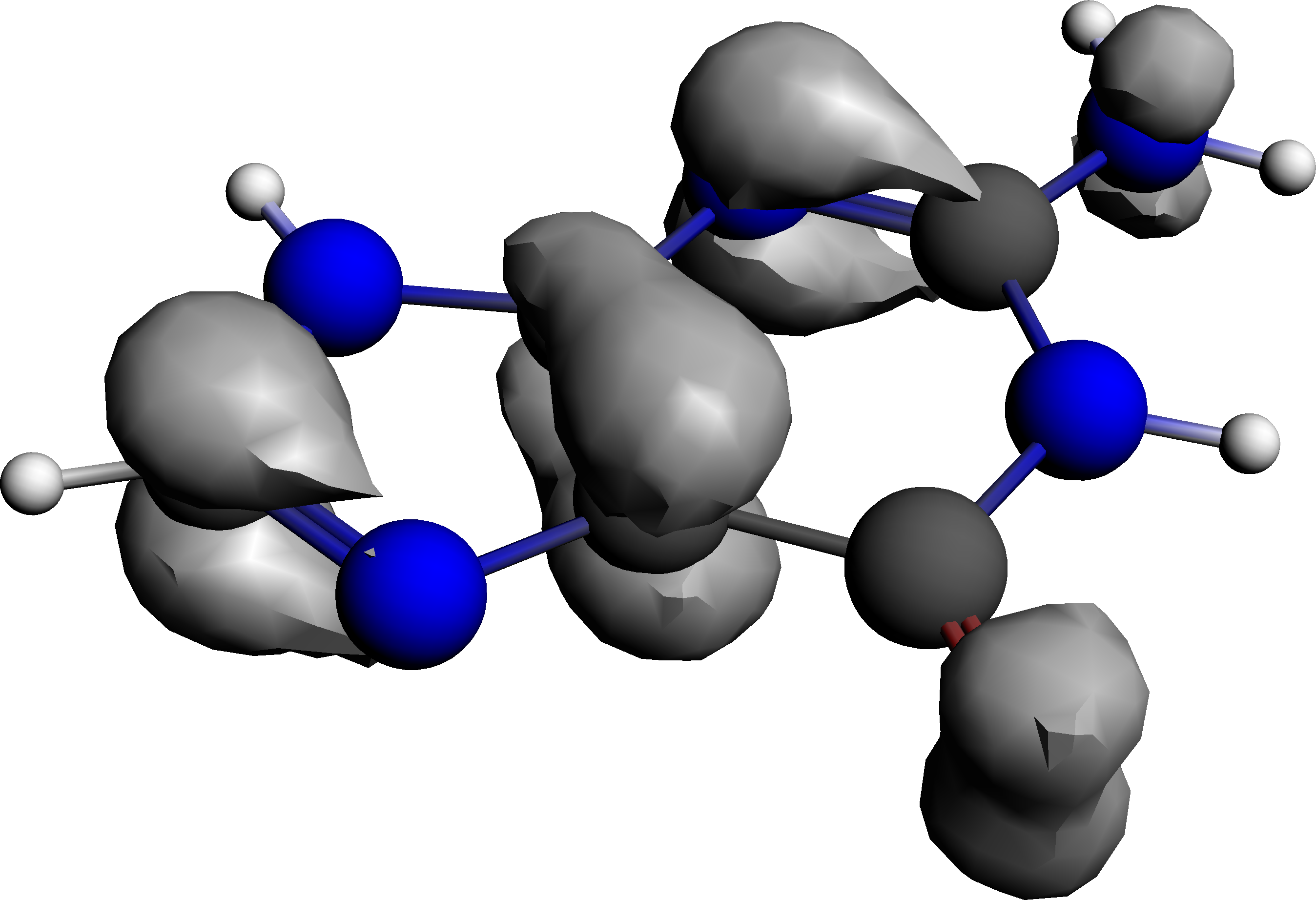}\\
(b)\\[1.5ex]
\includegraphics[width=8cm]{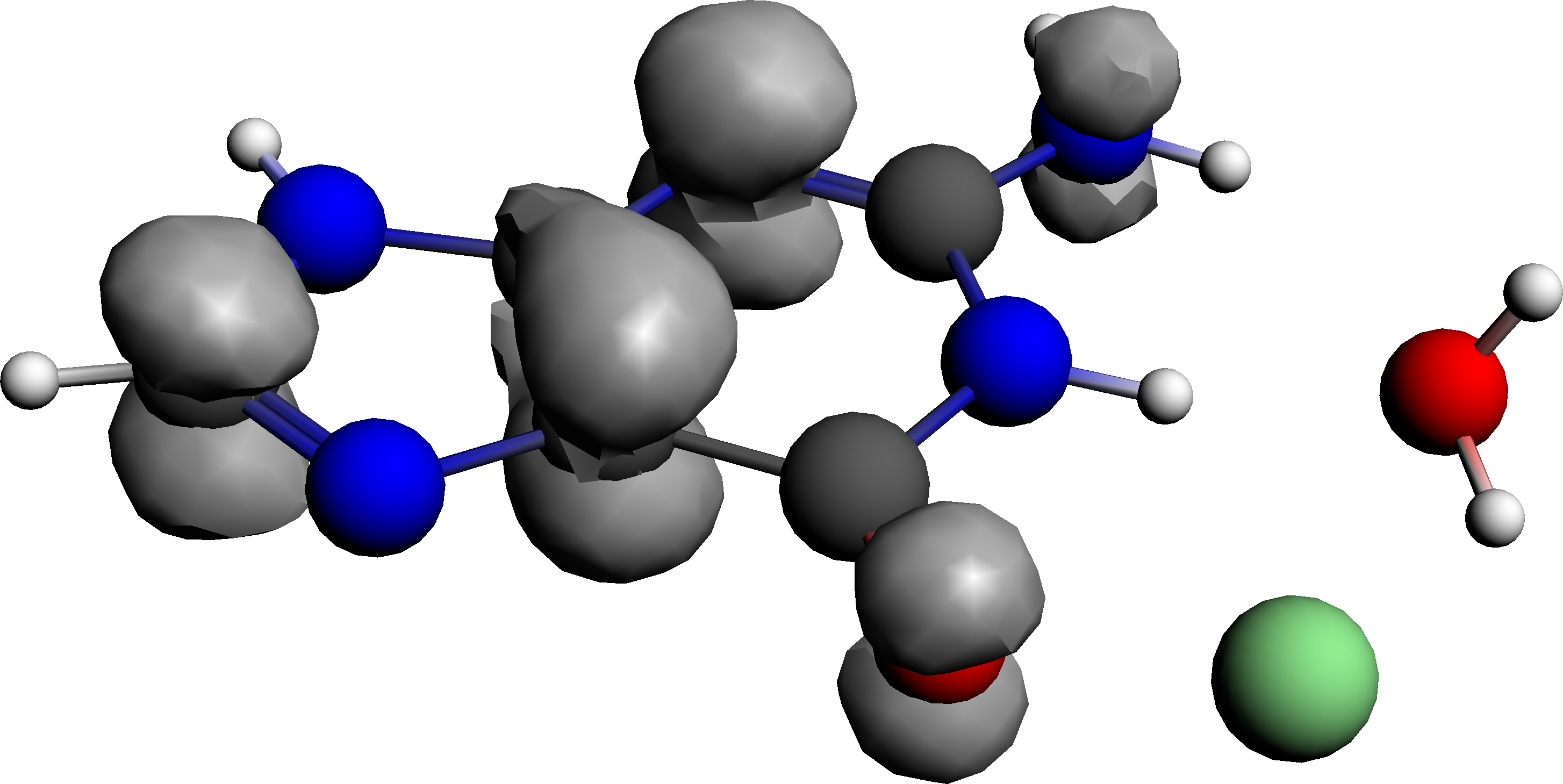}\\
(c)\\[1.5ex]
\includegraphics[width=8cm]{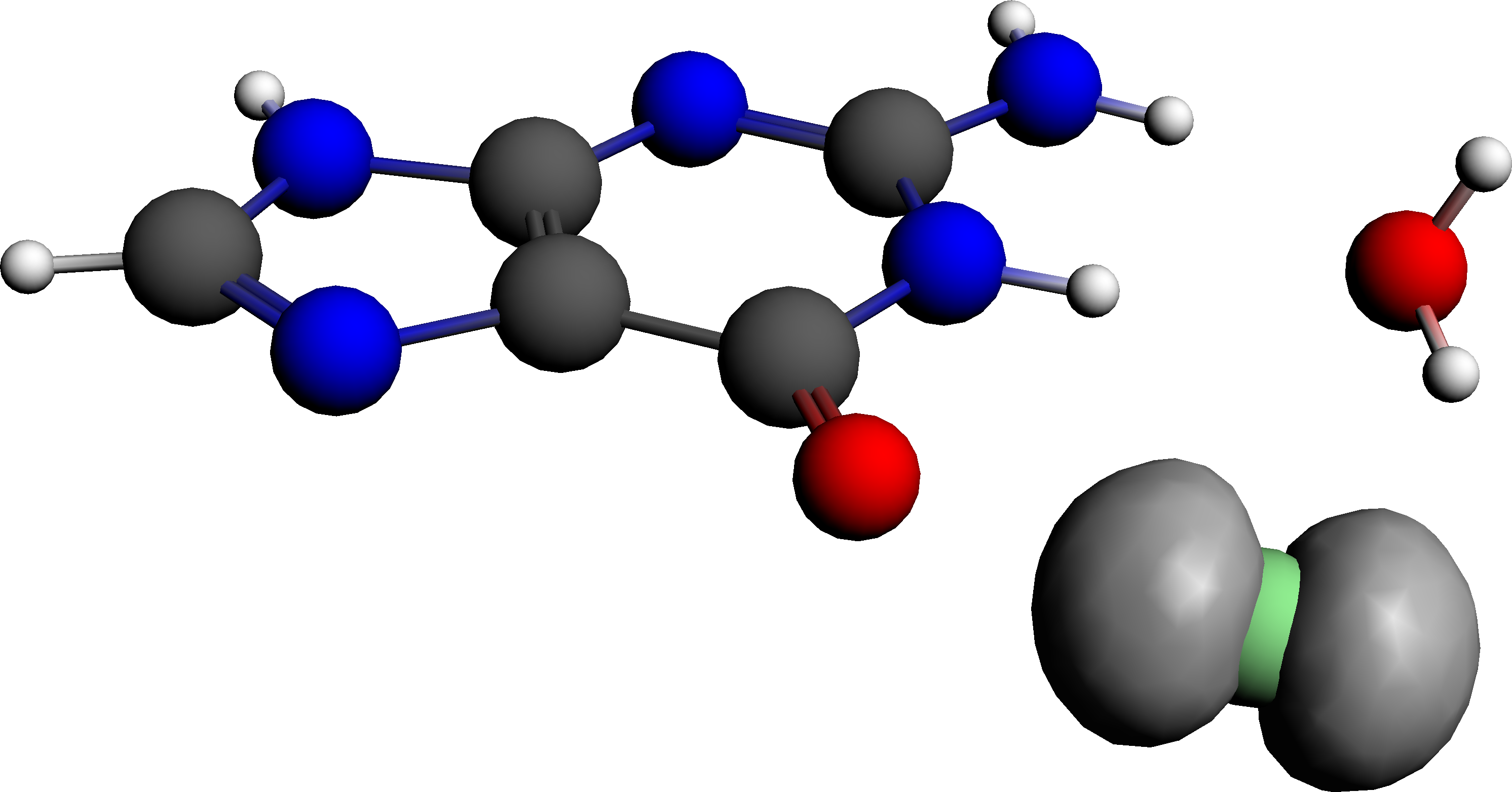}
\end{tabular}
\end{center}
\caption{\label{fig345} Spin density isosurface ($\rho^{\alpha}(\br)-\rho^{\beta}(\br)=0.005$) of guanine radical cation with one water molecule and one chlorine ion. Carbons, hydrogens, nitrogens, oxygens and chlorines are shown in gray, white, blue, red, and green, respectively. Inset (a) KS-DFT/M06-2X/TZP, (b) FDE/M06-2X/TZP, (c) KS-DFT/M06-2X/TZP (color online).}
\end{figure}

In contrast to FDE, the KS-DFT treatment of the system containing Cl$^-$ shows significant decrease of hfcc values for all atoms in the guanine moiety. This suggests that the KS-DFT spin density distribution for the ionic pair guanine-chlorine is very different from that in FDE and also from the one detected by the EPR experiments. Spin density population analysis in Figure \ref{fig345} shows that, contrary to the KS-DFT treatment, no inter-subsystem spin transfer occurs with FDE. I.e.\, FDE localizes the spin density entirely on the guanine radical cation. This density localization abilities of FDE is in line with previous studies \cite{solo2012,pava2011b}. Noteworthy, due to this localization ability, FDE effectively counteracts the self-interaction error which would necessarily enter the KS-DFT description.

The KS-DFT electron density distribution obtained for ionic pair guanine-chlorine [inset (c) in Figure \ref{fig345}] is well understood taking into account the corresponding ionization potential (IP) and electron affinity (EA) of the ions. The \textit{in vacuo} values of IP of chlorine ion and EA of guanine radical cation are 3.6 eV and 8.2 eV, respectively. Thus, even including the stabilization provided by the attractive Coulomb interaction, it is expected that electron density would readily flow from chlorine ion onto guanine radical cation resulting in chlorine atom and guanine molecule. This does not happen in the crystal. It is known, however, that similar to solvation, the ionic crystal environment considerably affects both IP an EA values of ions. For example, the IP of solvated chlorine ion is reported to be 8.33 eV \cite{gore1998}. Presumably, within the guanine crystal environment the IP of chlorine ion is considerably higher than 3.6 eV as well and, in turn, EA of guanine radical cation might be largely reduced. In cluster calculations carried out with regular KS-DFT, this effect may facilitate in correctly reproducing the electron density distrubution in the crystal. However, this is not the case in our calculations and, thus, KS-DFT is unfortunately not applicable here.

The above analysis demonstrates two important points. First, the bare guanine/water/chlorine model does not satisfactorily capture the embedding effects due to the crystal environment, as the predicted hyperfine coupling constants with this small model system depend strongly on the particular positions of the model constituents. A computational result in agreement with the experiment obtained with such a minimal model system should, therefore, be considered as both biased and coincidental. Second, KS-DFT is not able to reproduce the correct spin density distribution of the ionic pair guanine-chlorine within the crystal environment when small cluster model systems are considered. Keeping both of these points in mind, we proceed with calculations of isotropic hfccs for the larger, 7G and 36G model systems and exclusively using the FDE approach.

\subsection{Embedding effects evaluated on large QM models}

Our goal in this section is to characterize the embedding effects that are due to the molecules surrounding the bare (deprotonated) guanine. We do so by considering systems of increasing size: the 7G and 36G models. The results of corresponding FDE hfccs calculations are collected in the Table \ref{t5}.

\begin{table}
\begin{center}
\begin{tabular}{lrrrrr}
         &           & \multicolumn{2}{c}{{\bf 7G}} & \multicolumn{2}{c}{{\bf 36G}}\\
\hline
Atom & Ref.\ \cite{close1985}$^a$& TZP&TZ2P & TZP & TZ2P\\
                 N3                 & 16.8  & 11.3  & 11.4   & 15.4   & 15.5 \\
                 N10                & 10.1  &  7.4  &  7.5   &  9.5   &  9.6 \\
                 H$^{\prime}$       & 12.1  &-10.0  & -9.8   &-12.3   &-12.1 \\
                 H$^{\prime\prime}$ & 12.1  &-11.0  &-10.4   &-13.4   &-12.7 \\
                 H                  & 11.1  &-17.9  &-17.8   &-17.1   &-16.7
\end{tabular}
\end{center}
\caption{\label{t5} FDE M06-2X isotropic hfccs (MHz) of the 7G and 36G model systems of guanine radical cation in the crystal slab. Calculations carried out with TZP and TZ2P basis sets and TZP embedding density. $^a$ hfcc value for H atom in the last row is obtained from Ref.\ \cite{close1987}.}
\end{table}

Comparison between 7G and 36G models (Table \ref{t5}) shows an improvement of hfccs with increasing model size which rises a question whether the obtained results are converged with respect to the system size. To answer it, we embed the three model systems: bare guanine, 7G and 36G into a larger crystal pocket modeled via point charges which are obtained from M06-2X/TZP calculations on isolated protonated guanine cation and water molecule with the chlorine ion charge being equal to $-1$. The total system size considered is 15,119 atoms. All three models are treated at the FDE/M06-2X/TZP level of theory. The results are shown in the Table \ref{t6}.

\begin{table}
\begin{center}
\begin{tabular}{lrrrr}
Atom & Ref.\ \cite{close1985}$^a$& {\bf bare+MM}&{ \bf 7G+MM}&{\bf 36G+MM}\\
\hline
N3                             & 16.8  & 13.2  & 16.9  & 15.6 \\
N10                            & 10.1  &  9.2  & 10.4  &  9.2 \\
H$^{\prime}$                   & 12.1  &-12.1  &-13.2  &-11.9 \\
H$^{\prime\prime}$             & 12.1  &-13.0  &-14.4  &-13.0 \\
H                              & 11.1  &-16.8  &-16.8  &-17.1
\end{tabular}
\end{center}
\caption{\label{t6}FDE M06-2X isotropic hfccs (MHz) of bare guanine radical cation and 7G and 36G model systems embedded into a MM crystal pocket summing up to a total of 15,119 atoms in all cases. Calculations carried out with a TZP basis set. $^a$ hfcc value for H atom in the last row is obtained from Ref.\ \cite{close1987}.}
\end{table}

Table \ref{t6} reveals that the embedding effects are not entirely captured by the MM charges alone. For example, the hfcc values for the bare+MM system do not compare as well as the ones from 7G+MM and 36G+MM systems to the experimental values. The full QM embedding provided by FDE for the 7G and 36G systems features a better agreement with the experimental values and indicates that the inclusion of the soft electronic densities (as opposed to point charges) is key to an accurate accounting of embedding effects. This observation is in line with other studies that compared FDE embedding with an MM embedding \cite{jaco2006}. An \textit{ad-hoc} parametrization of the MM charges may improve the performance of the MM-only embedding model. However, we have not tested this hypothesis in the present work. In addition, the hfcc values for the 36G model system change only slightly with the inclusion of the MM charges. This shows that the 36G model already provides a large enough environment for this calculation.

\section{Conclusions\label{sec:conc}}

Several theoretical studies on guanine radical cation hyperfine couplings were carried out previously \cite{wetm1998,adhi2006,wang2012}. Similarly to what reported here, these studies were able to confirm the assignment of the EPR spectra to the deprotonated guanine. However, a comprehensive assessment of the crystal embedding effects on the calculated hfccs has never been carried out. We have shown that when a minimal environment model (Figure \ref{fig2}) is employed, the values of hfccs depend strongly on the chosen position of counter-ion and water molecule. We identified two non-equivalent minimal environment model systems. One reproduces the experimental values and one does not. This indicates that embedding effects play an important role for this system and that error cancellation is responsible for the fortuitous match with the experiment for one of the two small model systems. We also demonstrated that regular KS-DFT applied to cluster model systems of this ionic crystal is unable to provide reliable electron densities because of artificial electron density flow from chlorine ion to the guanine radical cation caused by the employment of finite cluster models as opposed to a crystal extending infinitely in all directions. 

Because the bare guanine model, with/without water or chlorine ion, is not able to capture the effects of the long-ranged Madelung field of the crystal,  it is not surprising that larger model systems, such as the 7G and 36G models, compare better to experimental hfcc values. The calculated hfccs collected in Table \ref{t1} for the bare guanine radical cation and in Tables \ref{t5} for 7G and 36G models show that the values for bare guanine and 7G models give quantitatively similar results with 36G model. The lack of any improvement going from the bare guanine to the 7G model can be qualitatively explained by the fact that despite its relatively large size (146 atoms) the 7G model is not able to fully capture the embedding effects due to the crystal environment.

Table \ref{t6} provides an important piece of information: {\it e.g.\ }the embedding effects are successfully captured by the QM/MM treatment. However, the closer the MM part is to the QM region of interest (in our case the deprotonated guanine) the less accurately the embedding effects are described by the simulations (as inferred by the comparison to the experiment). Even though in this work embedding effects are largely due to long-ranged electrostatics, the trend experienced by the calculated hfcc values with increasing QM region (bare+MM to 36G+MM in the table) indicates that a large QM environment surrounding the region of importance plays a non-negligible role and that FDE successfully captures these short-ranged effects that are missing in a MM-only embedding. The rather small difference between hfcc values for 7G+MM and 36G+MM models indicates that a failure to capture symmetry of the crystal as a result of treating not large enough model at QM level (7G) can be cured at QM/MM level (7G+MM). The very similar 36G and 36G+MM hfccs show convergence of the results with respect to the cluster size.

An even more accurate description of the considered system is feasible. For example, one may perform \textit{ab-initio} molecular dynamics simulations and average the hfccs over dynamical trajectories as done in Ref.\ \cite{neug2005f}. Such a calculation is, however, computationally very demanding for the system considered. Here, we followed a more computationally economical way and obtained data that provide unambigous support of the experimental observations.
\section*{Acknowledgements}
We thank the NSF-XSEDE program (award TG-CHE120105) for computational resources, and the office of the Dean of FASN of Rutgers-Newark for startup funds.
\section*{Supporting information available:}
Results of FDE calculations employing the PBE0 functional of hfccs for the 7G and 36G models as well as the cartesian coordinates of atoms of all model systems studied are available in the supporting information. This material is available free of charge via the Internet at http://pubs.acs.org.
%

\providecommand{\url}[1]{\texttt{#1}}
\providecommand{\urlprefix}{}
\providecommand{\foreignlanguage}[2]{#2}
\providecommand{\Capitalize}[1]{\uppercase{#1}}
\providecommand{\capitalize}[1]{\expandafter\Capitalize#1}
\providecommand{\bibliographycite}[1]{\cite{#1}}
\providecommand{\bbland}{and}
\providecommand{\bblchap}{chap.}
\providecommand{\bblchapter}{chapter}
\providecommand{\bbletal}{et~al.}
\providecommand{\bbleditors}{editors}
\providecommand{\bbleds}{eds.}
\providecommand{\bbleditor}{editor}
\providecommand{\bbled}{ed.}
\providecommand{\bbledition}{edition}
\providecommand{\bbledn}{ed.}
\providecommand{\bbleidp}{page}
\providecommand{\bbleidpp}{pages}
\providecommand{\bblerratum}{erratum}
\providecommand{\bblin}{in}
\providecommand{\bblmthesis}{Master's thesis}
\providecommand{\bblno}{no.}
\providecommand{\bblnumber}{number}
\providecommand{\bblof}{of}
\providecommand{\bblpage}{page}
\providecommand{\bblpages}{pages}
\providecommand{\bblp}{p}
\providecommand{\bblphdthesis}{Ph.D. thesis}
\providecommand{\bblpp}{pp}
\providecommand{\bbltechrep}{Tech. Rep.}
\providecommand{\bbltechreport}{Technical Report}
\providecommand{\bblvolume}{volume}
\providecommand{\bblvol}{Vol.}
\providecommand{\bbljan}{January}
\providecommand{\bblfeb}{February}
\providecommand{\bblmar}{March}
\providecommand{\bblapr}{April}
\providecommand{\bblmay}{May}
\providecommand{\bbljun}{June}
\providecommand{\bbljul}{July}
\providecommand{\bblaug}{August}
\providecommand{\bblsep}{September}
\providecommand{\bbloct}{October}
\providecommand{\bblnov}{November}
\providecommand{\bbldec}{December}
\providecommand{\bblfirst}{First}
\providecommand{\bblfirsto}{1st}
\providecommand{\bblsecond}{Second}
\providecommand{\bblsecondo}{2nd}
\providecommand{\bblthird}{Third}
\providecommand{\bblthirdo}{3rd}
\providecommand{\bblfourth}{Fourth}
\providecommand{\bblfourtho}{4th}
\providecommand{\bblfifth}{Fifth}
\providecommand{\bblfiftho}{5th}
\providecommand{\bblst}{st}
\providecommand{\bblnd}{nd}
\providecommand{\bblrd}{rd}
\providecommand{\bblth}{th}

\end{document}